\documentclass{article}
\usepackage[top=1.5in, bottom=1.5in, left=1.2in, right=1.2in]{geometry}
\usepackage{authblk}
\usepackage{amssymb, amsmath,framed}
\usepackage{graphicx}
\usepackage[round]{natbib}

\usepackage{bm}
\usepackage[colorlinks=true,linkcolor=blue,urlcolor=blue,citecolor=blue,anchorcolor=blue]{hyperref}
\expandafter\ifx\csname package@font\endcsname\relax\else
 \expandafter\expandafter
 \expandafter\usepackage
 \expandafter\expandafter
 \expandafter{\csname package@font\endcsname}
\fi
\def\bq{\begin{equation}}
\def\eq{\end{equation}}
\def\bqy{\begin{eqnarray}}
\def\eqy{\end{eqnarray}}

\makeatother

\begin{document}

\title{What's in a name: The etymology of astrobiology}

\author{Manasvi Lingam\thanks{Electronic address: \texttt{mlingam@fit.edu}}\,\,}

\affil{Department of Aerospace, Physics and Space Science, Florida Institute of Technology, Melbourne FL 32901, USA}
\affil{Institute for Theory and Computation, Harvard University, Cambridge MA 02138, USA}
\author{Abraham Loeb}
\affil{Institute for Theory and Computation, Harvard University, Cambridge MA 02138, USA}

\date{}

\maketitle

\begin{abstract}
Astrobiology has been gaining increasing scientific prominence and public attention as the search for life beyond Earth continues to make significant headway on multiple fronts. In view of these recent developments, the fascinating and dynamic etymology of astrobiology is elucidated, and thus shown to encompass a plethora of vivid characters drawn from different continents, religions, ideologies and centuries. \\
\end{abstract}

\section{Introduction}
Notwithstanding the fact that Shakespeare's celebrated quote from Act II, Scene II of \emph{Romeo and Juliet} (``What's in a name?'') has been, at times, (mis)interpreted as signifying the arbitrariness of nomenclature, there is no doubting the reality that names and words do play a major role in human culture \citep{Lo47,Cas44,Jes09,Alt14}; this aspect has been appreciated since at least the days of Confucius, who argued in favour of ``rectifying'' names \citep[pg. 57]{Gard14}, and Plato, who wrote about the ``correctness of names'' in his \emph{Cratylus} \citep{Sed18}. At the same time, however, it should be noted that the precise semantic status of names and words continues to attract much debate and controversy \citep{Aus75,Str16,SC19,Spe19}. 

The transdisciplinary endeavour of astrobiology has swiftly gained prominence in the realm of academia as well as the public consciousness in the past few decades \citep{RDB12,Vak13,SJD18,CPMD,Craw18}. With the increasing scientific and media buzz surrounding astrobiology, even as we continue to make considerable progress toward resolving this question, it seems a worthwhile endeavour to step back and reflect on the origins of this word. In doing so, we shall encounter a bevy of forgotten actors, some of whom were pioneers of this variegated field, and consequently gain a deeper historical awareness of how the myriad meanings underpinning astrobiology have evolved and morphed over the ages. 

It is essential to emphasize at the outset that this pr\'ecis is \emph{not} meant to serve as a comprehensive historical overview of the field; the reader may consult \citet{Di82,Cr088,Mic07,CD13,SJD18} for the same. Shortly after the Copernican revolution, which is often (and perhaps somewhat erroneously) credited with displacing the Earth from its privileged position at the centre of the physical universe \citep{Ku57,Gin85,Hu17,Ca18}, the philosopher Giordano Bruno (1548-1600) (in)famously averred in \emph{De l'infinito, universo e mondi} that there exist countless stars hosting planets teeming with multifarious lifeforms, both like and very unlike those found on Earth \citep[pg. 144]{Bo14}; see also \citet[pp. 19-20]{Wi03}. 

The postulation of extraterrestrial life and the associated notion that the Earth is not located at the centre of the biological universe has a rich (albeit neglected) history prior to the advent of Bruno, encompassing personages as diverse as Anaximander, Democritus, Epicurus, Lucretius, Muhammad al-Baqir, Moses Maimonides, Fakhr al-Din al-Razi, Albertus Magnus, Hasdai Crescas, Nicholas of Cusa, and William Vorilong \citep{Tip81,Cro97,Bra06,Sch10,We14}. Aside from this group of scholars, many religious traditions and myths from all over the world had also posited that the Earth is not alone in hosting life \citep{Het93,SS00,Rug05,Naz09,We14}; in particular, most of the major surviving South Asia religions fall under this category. In all of the above instances, while the word ``astrobiology'' may not have explicitly appeared as such, it is apparent that these writings constitute the forerunners of astrobiology in several respects.

Before we forge ahead, it is valuable to recapitulate the main objectives of astrobiology \citep{MW99,CH05,Lun05,SDW16}, which can be roughly broken down into three major questions.\footnote{\url{https://www.nasa.gov/50th/50th_magazine/astrobiology.html}}
\begin{enumerate}
    \item How does life begin and evolve? (Where did we come from?) 
    \item Does life exists elsewhere in the universe? (Are we alone?)
    \item What is life’s future on earth and beyond? (Whither are we going?)
\end{enumerate}
In what follows, we will mostly direct our attention toward the second question (Question \#2). Our chief rationale is that the majority of early references that explicitly invoked the word ``astrobiology'' did so in the context of the second question, as we shall see hereafter. 

It should, however, be recognized that both the first and third questions have a long and rich history in their own right, even though the pertinent early references in these areas did not use the word ``astrobiology'' as such. For a history of the origin and evolution of life (Question \#1) - a multi-disciplinary endeavour in its own right that draws upon fields as diverse as geochemistry and information theory - the reader may consult \citet{Sch92,CM95,Knoll15,Lu16,DW16,MB19}. The future of life on Earth (Question \#3) has garnered relatively less attention,\footnote{It could be argued that this domain, which synthesizes domains as diverse as astronomy, engineering and ecology, has received short shrift on the whole.} despite the fact that the present-day dynamic environment of Earth is exerting, and will continue to exert, a profound effect on the future of life on Earth. Reviews and analyses of Question \#3 can be found in \citet{Dys79,CK92,Cir12,Vid14,WT15}.

At this stage, it is also necessary to clarify the difference between ``exobiology'' and ``astrobiology'', as there has been a tendency in some quarters to conflate and/or equate these two words. Exobiology forms the core of Question \#2, as it primarily deals with the question of gauging the ``cosmic distribution of life'', to use the phrase invoked by the Nobel laureate Joshua Lederberg in his seminal work \citep[pg. 393]{Led60}; this paper is often credited with playing a noteworthy role in driving NASA's exobiology programs in the early years \citep{CH05,Di09}. Hence, when viewed in this spirit, exobiology may be perceived as a sub-discipline of astrobiology; from such a perspective, it would be erroneous to conclude that astrobiology was merely an offshoot of exobiology. As we will mostly tackle astrobiology from the standpoint of Question \#2 henceforth, we shall be implicitly operating in the domain of exobiology.

\section{Tracing the etymology of astrobiology}
The goal of this treatise is to trace the major milestones in the etymology of ``astrobiology''. Needless to say, this survey is by no means exhaustive, as there could be any number of works that have been overlooked due to the limited access to early publications in conjunction with the attendant difficulties in perusing non-English and non-European references. As a result, we will focus on four distinct cases, corresponding to the potential first appearance of this word in: (i) a scientific monograph, (ii) a publication in an established scientific journal, (iii) a ``popular science'' periodical or an academic journal not devoted to the sciences, and (iv) a work of fiction or outside the domain of science. Before embarking on this journey, NASA's role in developing astrobiology during its nascency is worth highlighting. Shortly after its inception in 1958, NASA commissioned a variety of far-sighted projects ranging from experiments in space biology to the construction of life-detection instruments for ambitious future missions to Mars \citep{DS05,Di09}.

\subsection{Scientific monographs}
Our voyage into the past begins with (i). Perhaps the very first \emph{scientific} book bearing the title of \emph{Astrobiology} (Russian title: \emph{Astrobiologiya}) was published by the Belarusian polymath Gavriil Adrianovich Tikhov (1875-1960) in 1953 \citep{Tik53}.\footnote{Although the following books did not explicitly incorporate the word ``astrobiology'' in the title, notable examples from the same period and much earlier that tackled this theme include \citet{deF67,Whe67,Fla71,Wal03,Jo40,Str53,Sh58,Dole}.} On account of having carried out his research almost exclusively in the Soviet Union, Tikhov's contributions to astrobiology remained mostly forgotten for decades. However, as illustrated by recent studies \citep{OT05,Bri13}, Tikhov undertook a number of visionary projects in astrobotany and astrobiology, aside from pursuing conventional topics such as variable stars, comets and the Sun. Tikhov's research in astrobiology encompassed measurements of Earthshine \citep{Tik14}, analyses of plant physiology in extreme physicochemical conditions, and experiments to characterize the spectral properties of plants and consequently assess how their analogs could be detected on Mars \citep{Tik55}. We find echoes of these areas in current astrobiology, ranging from studies of extremophiles \citep{RM01,MAB19} to next-generation searches for the ``red edge'' of vegetation (induced by the presence of chlorophylls) on exoplanets \citep{STSF}; in principle, similar signatures, albeit of the technological kind, might also arise because of artificial photosynthesis \citep{LL17}.

However, even prior to Tikhov's tome, there was at least one other book that employed the word ``astrobiology'' in the title, albeit not in the modern scientific sense of this word. The monograph in question is \emph{La pens\'ee de l'Asie et l'astrobiologie} (1938) by the French philosopher and historian Ren\'e Berthelot (1872-1960). In this work on anthropology, astrobiology signified the stage of human development in which human societies subscribed to animistic or vitalistic interpretations of natural phenomena in parallel with a certain degree of astronomical knowledge, and a belief that the latter shaped terrestrial phenomena \citep{Ber38,Lem10}. Although Berthelot's conception of astrobiology has  altogether fallen out of use in the 21st century, it was employed \emph{in hoc sensu} by French intellectuals up to the late 20th century \citep[pg. 4]{Chr19}.

\subsection{Scientific journals}
Next, we turn our attention to (ii). It is widely supposed that Lawrence J. Lafleur (1907-1966) - a philosopher at Brooklyn College who was better known for his translations of Ren\'e Descartes' works - authored the first peer-reviewed scientific publication entitled \emph{Astrobiology} \citep{Blum,Ba05,CH05}, in which he defined astrobiology as, ``the consideration of life in the universe elsewhere than on earth'' \citep[pg. 333]{Laf41}. Although the modern interpretation and scope of astrobiology are broader, because it also encompasses the origin and evolution of life on our planet, Lafleur honed in on many of the key topics and goals of this field; in point of fact, his definition and analysis were not far removed from the much better known exposition of exobiology by Joshua Lederberg in 1960, nearly two decades later \citep{Led60}. 

In this unusually prescient publication, Lafleur singled out many of the basic requirements for habitability, as seen from the following quote \citep[pp. 333-334]{Laf41}:
\begin{quote}
One of the important considerations is the chemical constitution, involving fairly high proportions of carbon, oxygen, nitrogen and hydrogen, together with smaller quantities of a large number of elements, and their existence in such proportions that the compounds found on earth could exist, particularly water. Other requirements include: a temperature like that on earth; a pressure at the surface not too dissimilar to conditions here \dots a source of light energy adequate to keep plants alive \dots    
\end{quote}
In the same article, Lafleur outlined the possibility of detecting signatures of intelligent extraterrestrial life by searching for ``interstellar communication'' \citep[pp. 338-339]{Laf41}, thus anticipating, to an extent, the seminal paper by \citet{CM59} that is conventionally regarded as having initiated the Search for Extraterrestrial Intelligence (SETI) (see \citealt{Tar01}). At the risk of digressing, we note that the notion of employing electromagnetic signals for communication was espoused in a scientific publication as early as 1931. In the journal \emph{Nature}, Ernest William Barnes (1874-1953), the Bishop of Birmingham, opined that \citep[pg. 722]{Bar31}:
\begin{quote}
As I have already indicated, I have no doubt that there are many other inhabited worlds, and that on some of them beings exist who are immeasurably beyond our mental level. We would be rash to deny that they can use radiation so penetrating as to convey messages to the Earth. Probably such messages now come. When they are first made intelligible a new era in the history of humanity will begin.
\end{quote}

\subsection{Popular-science and non-scientific journals}
Even though Lafleur has been credited by some sources as the first person to coin the word ``astrobiology'' \citep{Blum}, recent scholarship by \citet{Bri12} has revealed that the Polish scientist and engineer Ary J. Sternfeld (1905-1980) introduced this term in the French popular science magazine \emph{La Nature} in 1935 \citep{Stern}, thereby exemplifying (iii). His article was filled with a number of prescient musings, of which one of the most perspicacious was the prediction that Titan probably possessed an atmosphere; another that stands out in the context of this article is his apposite definition of astrobiology \citep{Bri12}:
\begin{quote}
The development of both the natural and astronomical sciences has led to the birth of a new science whose main objective is to assess the habitability of the other worlds, this science is called astrobiology. 
\end{quote}
Sternfeld's peripatetic existence led him from the small town of Sieradz, Poland (his place of birth) to Paris and thence to Moscow. Despite facing numerous hardships as a Jewish person navigating the turbulent politics of 20th century Europe, his breakthroughs in astronautics have been gaining belated recognition. In this realm, Sternfeld is known to have presented the mathematical details underpinning bi-elliptic transfer \citep{Ster} - an intricate maneuver that requires lower delta-v in comparison to the famous Hohmann transfer under certain conditions \citep[pg. 82]{DM19} - and coined the term ``cosmonautics'' in his book \emph{Initiation \'a la Cosmonautique} \citep{Iv03}. While the above quote makes it clear that Sternfeld's view of astrobiology is eerily reminiscent of its modern interpretation, there were preceding works that espoused very different notions of what defined and comprised astrobiology. 

In particular, a few publications from the first few decades of the 20th century suggested that astrobiology comprised the biological rhythms and other effects engendered by (lunar) tides. For instance, \citet[pg. 664]{Bru08} described the observed correlation between the lunar and reproductive cycles for certain marine species, most notably \emph{Eunice viridis}, as an ``\emph{enigma astrobiologico}'' (astrobiological enigma). Gustavo Brunelli (1881-1960), the Italian biologist who authored this article, also wrote an altogether forgotten book in 1933 centered around the origin and evolution of life on Earth;\footnote{\url{http://www.treccani.it/enciclopedia/gustavo-brunelli_(Dizionario-Biografico)}} in this early treatise, among other insights, he underscored the significance of condensation reactions for synthesizing proteins from amino acids \citep[pg. 177]{Bru33}. Another publication that emphasized the role of tides in an astrobiological context was by the French writer Maurice Privat (1889-1949). In \citet{Pri36}, he posited the importance of ``\emph{le rythme lunaire}'' (lunar rhythms) for comprehending the biological characteristics of humans and other species. Although similar biological implications of tides have been explored in modern astrobiology, not surprisingly, they constitute a minuscule fraction of this diverse field \citep{LL18}. 

Before moving ahead, there is one other interesting reference that merits a mention. In the journal \emph{Archiv f{\"u}r systematische Philosophie}, the German doctor and inventor Ferdinand Maack (1861-1930) - credited by some sources as the pioneer of \emph{Raumschach}, i.e., a three-dimensional version of chess \citep[pg. 419]{HW96} - touched upon the subject of astrobiology in 1918. In fact, \emph{contra} the majority of contemporaneous publications, Maack's notion of \emph{Astro-Biologie} (astro-biology) was arguably quite modern, as evinced by the rough translation of \citet[pg. 45]{Ma18}:
\begin{quote}
We transition from mainly mechanical and physical questions to biological and psychological ones. The fourth group of problems deals with astro-biology and astro-psychology. \\
How did the first life on Earth originate? Through spontaneous generation? Cosmic panspermia? How does the human race and life on our planet end? Is only the Earth inhabited? Or do higher, human-like, beings dwell on other worlds?
\end{quote}
In posing the above questions, Maack echoes some of the ideas propounded by Bruno and his forebears, and it is therefore apparent that he viewed astrobiology as the science of extraterrestrial life, even if no explicit definition was furnished as such.

\subsection{Non-scientific monographs and works of fiction}

The last step on this winding road is (iv). The history of science is replete with words and phrases that originated in non-technical publications, arguably most notably in works of fiction, with the quintessential example being ``quarks'' in the realm of particle physics - a word that was adopted from James Joyce's classic \emph{Finnegans Wake} by Murray Gell-Mann \citep[pg. 180]{GM94}. There are compelling grounds for believing that ``astrobiology'' may constitute another such striking exemplar. A meticulous search, implemented by means of utilizing the Google Books and Google Scholar search engines among others,\footnote{\url{https://books.google.com/} and \url{https://scholar.google.com/}} for this word yields only a handful of bona fide results. The majority of these sources are highly inchoate, with the rest of them displaying a predominantly theological bent.

The most coherent quotes in the latter category are interspersed across the writings of Cyrus Teed (1839-1908), a charismatic eclectic physician turned religious leader, who founded the sect known as the ``Koreshan Unity'' after experiencing a vision from ``The Divine Motherhood'' in 1869 \citep{Mill}. One of the most unusual aspects of this group was their belief in a unique variant of the Hollow Earth theory, namely, an inside-out cosmology wherein humans, the biosphere, and the celestial objects inhabited the \emph{interior} of a hollow shell \citep{Teed}.\footnote{\url{https://www.lockhaven.edu/~dsimanek/hollow/morrow.htm}} The Koreshan Unity, which was based on the principles of communal living and eventually settled in Florida, reached its peak membership of $250$ followers belonging to sundry backgrounds in 1908, and thereafter declined with the last member (Hedwig Michel) joining in 1940 after fleeing Nazi Germany \citep{Mill}.

In \emph{The Cellular Cosmogony} (1898), widely regarded as the summation of Teed's thought, astrobiology was invoked in this outr\'e passage \cite[pg. 22]{Teed}:
\begin{quote}
\dots and further, that when the Lord was visibly manifest to the outer world, his inner and spiritual life was visible to the spiritual world as the astrobiological center of that sphere, and beside him there was no God.
\end{quote}
In the June 8, 1900 issue of the magazine \emph{The Flaming Sword}, published under the aegis of Koresh (viz., Teed),\footnote{It must, however, be noted that the actual author of this work remains anonymous, i.e., there is no concrete evidence that Teed wrote this article.} a comparatively down-to-earth and intriguing, albeit terse and therefore ambiguous, definition of astrobiology was promulgated \cite[pg. 4]{Koresh}:
\begin{quote}
Cosmogony includes the earth, sun, stars, planets, and all life--in a more narrow sense, the alchemico-organic system \dots If we use the term astrobiology, we would mean the stars and all life.
\end{quote}
This quote might very well be the first explicit semi-definition of astrobiology specified in an English-language publication, although its general tenor is manifestly removed from the current interpretation of what constitutes the domain of astrobiology. A few years thereafter, in the January 15, 1908 issue of the \emph{The Flaming Sword}, Teed conceived of astrobiology as the ``regulation of human affairs by the clock-work of the Cosmos'', and argued that the function of astrobiology was ``to determine career before it is brought to the birth or before conception'' \citep{Ko08}.

At approximately the same time that Cyrus Teed was publishing his religious writings, the French philosopher Henry Lagr\'esille (1860-19xx) authored \emph{Le Fonctionnisme Universel} in 1902, a book dealing primarily with metaphysics that was not well-received by contemporary reviewers \citep{Mor03,Rey}. In this work, he envisioned astrobiology as ``\emph{loi qualitative de l'\'energie}'' \citep[pg. 540]{Lag02}, i.e., as the ``qualitative law of energy'' \citep[pg. 4]{Chr19}.\footnote{In modern astrobiology, concepts such as free energy and thermodynamic disequilibrium play a vital role in constraining the origin and evolution of putative biospheres \citep{Hoe07,BBGR}.} In the only instance where \emph{astrobiologie} (astrobiology) explicitly appears \citep[pg. 541]{Lag02}, a rough translation of that passage is furnished below:
\begin{quote}
You can thus conceive that today’s astronomy only offers celestial mechanics, and therefore provides only the abstract framework of a more concrete science, in closer conformity with [living] beings - namely astrobiology - which, if it were possible, would adopt the essence of astrology, not unlike how chemistry superseded that of alchemy. For, once the link between these two fields has been established, the more or less obscure empirical laws, which had been unveiled by intuitive revelations, would find, to a certain extent, their rational explanations in this conception of a cosmic life awash with finalities and conscious forces.
\end{quote}
Among publications from this period, it should be noted in passing that the German art historian Willy Pastor (1867-1933) invoked \emph{astrobiologie} a few times in his five-act play \emph{Das Reich des Krystalls} (The Realm of the Crystals) published in 1901, but the references to this word are fleeting, vague and incoherent \citep[pp. 59-60, 71]{Pas01}.

Amidst this motley crowd of early references to astrobiology, one other work stands out to some extent, which was briefly noted in \citet{NVS15}. The work in question is \emph{Limanora: The Island of Progress}, a science-fiction novel written by Godfrey Sweven in 1903 \citep{Swe03}. \emph{Limanora} distinguishes itself from the multitude of contemporaneous allusions to astrobiology on two fronts: (i) this term is employed several times in the book, indicating that its usage was no fluke, and (ii) it represents a genuine piece of science fiction that was inspired, at least in part, by Jonathan Swift's \emph{Gulliver Travels} and the novels of H. G. Wells. It is worth noting that \emph{Limanora} evinces a distinct anti-racist and anti-colonial stance in certain respects, but the novel concomitantly reified (\emph{verdinglichte}) some of the prevalent colonialist and racist notions of its era; for example, it espoused a variant of Social Darwinism \citep[pp. 72-74]{Rie}. 

The author of this book was a rather interesting personage in his own right, owing to which we shall delve into this subject briefly. Godfrey Sweven was a pseudonym utilized by the Scottish-born New Zealand academic and administrator John Macmillan Brown (1845-1935). As a academic, Brown published articles and books in areas ranging from analyses of English literature to the cultures of the Pacific islands, but his scholarship in the latter area was ``regarded with scepticism and strongly criticised by Apirana Ngata and Te Rangi Hiroa'' as duly pointed out in \emph{Te Ara: The Encyclopedia of New Zealand}.\footnote{\url{https://teara.govt.nz/en/biographies/2b41/brown-john-macmillan}} Brown was also known for being an advocate of higher education for women; while he was based at Canterbury College, New Zealand, Brown purportedly played a key role in admitting Helen Connon. She subsequently became the first woman to receive a university degree with honours in the British Empire, and gained recognition as a gifted scholar and accomplished educator.

Returning to \emph{Limanora}, the natural question that arises is: Was the phraseology of astrobiology ``modern''? This is not a facile question to resolve, seeing as how an unambiguous definition of astrobiology is not spelt out in the novel. However, by undertaking a close reading of the text, the tentative answer is probably in the negative. More specifically, Brown appears to employ this term to signify something akin to present-day genetic engineering, as illustrated by the sentence following the usage of ``astrobiology'' \citep[pg. 309]{Swe03}:
\begin{quote}
Soon would they modify and improve the lavolan to fit the conditions of interstellar space, and the faleena, if not their own organs, for venturing far into the rarest ether. And then what reports, what pictures of the invisible universes would they bring before the eyes and the firlas of their fellow-islanders!
\end{quote}

\section{The future of astrobiology}
With our analysis concerning the etymology of astrobiology having drawn to a close, it seems apropos to end our discussion with a tour d'horizon of the past, present and future of this blossoming field. Over the past decade, we have discovered complex organic molecules within the interstellar medium and on objects within our Solar system such as meteorites, Mars and Enceladus \citep{SI16}, identified Earth-sized planets in the temperate zones of other stars and refined some basic requirements for habitability \citep{Per18,LL19}, designed generic prebiotic pathways that are reliant on the availability of ultraviolet radiation \citep{Su17}, furthered our understanding of how metals could catalyse protometabolic networks \citep{PX19}, unearthed animals that lack mitochondria altogether \citep{YAN20}, expanded the parameter space of the limits tolerated by (poly)extremophiles \citep{MAB19}, conceived novel methods of identifying biosignatures \emph{in situ} as well as via remote sensing \citep{SKY18,NHV18}, and taken our very first steps toward undertaking high-speed interstellar travel \citep{Pop17,WDK18}, to name a few. 

Yet, at the same time, it is equally essential to recognize that the preceding picture is deliberately rosy-hued. In many respects, we are very far from settling the question(s) of how, where and when life arose and evolved on our planet, to say nothing of other worlds. Hence, it is vital to avoid misconstruing our current rate of progress and thereby lulling ourselves into a state of false complacency. As Shakespeare wisely wrote in Act II, Scene III of \emph{All's Well That Ends Well}:
\begin{quote}
They say miracles are past; and we have our philosophical persons, to make modern and familiar, things supernatural and causeless. Hence is it that we make trifles of terrors, ensconcing ourselves into seeming knowledge, when we should submit ourselves to an unknown fear.
\end{quote}
However, in place of succumbing to ``an unknown fear'', as ostensibly espoused by Shakespeare's Lafeu, we can (and should) instead opt to temper our optimistic hopes and expectations with a judicious dose of caution and skepticism, thus taking equal delight in what we discover and comprehend, and what remains unseen and unknown.

If we look ahead to the future, even as humanity braces itself to confront a host of grave challenges and growing schisms \citep{BC08,Kol14,Kl15,Rees}, we may cautiously identify multiple reasons to be optimistic about the scientific future of astrobiology in the upcoming decade(s).\footnote{In the best-case scenario, astrobiology might even aid humankind in navigating the turbulent currents of the Anthropocene \citep{Frank}.} We will, in all likelihood, witness the launch of large ground- and space-based telescopes to commence the hunt for novel biosignatures \citep{FAD18}, life-detection missions to potentially inhabited worlds such as Europa and Mars \citep{NHV18}, experiments that push the boundaries of the long-term survival of biota in the harsh conditions of outer space \citep{MCR16}, theoretical and experimental breakthroughs in resolving the question of how life originated on our planet \citep{Walk,LL21}, advances in comprehending how Earth's biogeochemical cycles have coevolved over time and the ensuing implications for our biosphere \citep{KN17}, among many others. 

Hence, at the risk of retreading and extolling a hoary clich\'e, while the age-old question of ``Are we alone?'' has captivated humanity for millennia, it is not much of an exaggeration to contend that we might find ourselves situated on the cusp of a momentous era wherein we can hope to settle this question scientifically through the synthesis of experiments, observations and modelling. In light of the aforementioned putative future developments, a quotation from \emph{Limanora} - which comprises one of the earliest allusions to astrobiology, as stated previously - is strikingly vatic, albeit when viewed out of context \citep[pg. 309]{Swe03}:
\begin{quote}
For astrobiology they saw at a glance there was begun a new and lofty career.
\end{quote}

\section*{Acknowledgments}
We are grateful to Jean Schneider for pointing out a couple of valuable early references in the realm of astrobiology, Jeremy Riousset for generously sharing his feedback regarding a couple of translated passages, and the editor Rocco Mancinelli for his sagacious feedback. This work was supported in part by the Breakthrough Prize Foundation, Harvard University's Faculty of Arts and Sciences, and the Institute for Theory and Computation (ITC) at Harvard University.


\end{document}